\documentclass[10pt]{blois07}
\usepackage{graphicx}
\usepackage{cite,./mcite}

\begin{document}
\title{Saturation effects in elastic scattering  at the LHC }
\author{O.V. Selyugin$^1$\protect\footnote{\ \ talk presented at EDS07},
 J.-R. Cudell$^2$}
\institute{
$^1$selugin@theor.jinr.ru, BLTPh, JINR, Dubna\\
$^2$JR.Cudell@ulg.ac.be, IFPA, AGO Dept., Universit\'{e} de Li\`{e}ge}
\maketitle
\begin{abstract}

   The problems linked to the extraction of the basic parameters of
 the hadron elastic scattering amplitude at the LHC are explored.
  The impact of the Black Disk Limit (BDL)$-$ which constitutes a
new regime of the scattering processes
 $-$ on the determination of these values is examined.
\end{abstract}

\section{Introduction}

    The diffraction processes will occupy an important place
  in the experimental program at the LHC. Firstly,  we will need to know the
  luminosity and the total cross section with a high precision. Secondly,
  the diffraction processes
  will be directly explored at the LHC and will contribute to many different
  observable reactions.
 The planned analyses very clearly have problems from the theoretical view point.
For example, the definition of the differential cross sections of the elastic
proton-proton scattering, as presented in \cite{Royon06}
\begin{equation}
\frac{dN}{dt} = {\cal{L}} [\frac{4 \pi \alpha^2}{|t|^2} -
       \frac{\alpha \rho \sigma_{tot} e^{-b|t|/2}}{|t|}
      +\frac{\sigma_{tot}^2(1+\rho^2)e^{-b|t|} }{16 \pi}]
\end{equation}
does not contain the electromagnetic form factor and the
 Coulomb-hadron interference phase $\Phi_{CH}$. Such terms have to be included:
 all the corrections to $\phi_{CH}$ were calculated in
 \cite{SelyuginMP96}.    
More importantly, Eq. (1) is based on the assumption of an exponential behavior of the imaginary and real parts of the hadron scattering amplitude, which is at best an approximation.

Furthermore, the TOTEM experiment has announced the extraction  of
  $\sigma_{tot}$ from the experimental data, assuming a fixed value of $\rho(s,t=0)=0.15$.
  Indeed, the impact of $\rho$ on $\sigma_{tot}$ is connected with the term $(1+\rho^2)$,
 and is very small when $\rho$ is small. However the most important correlation
   between $\rho$ and $\sigma_{tot}$ enters the analysis
   through the Coulomb-hadron interference term, the size of which remains unknown
   if we do not know  the  normalization of $dN/dt$
   and the size and $t$-dependence  of $\rho(s,t)$ and $\phi_{CH}$.

In \cite{Royon06}, it was shown that there would be
large correlations between the value of $\rho$ and that of $\sigma_{tot}$.
However, these correlations and the error estimates were obtained
using an exponential behavior of the imaginary and real part of the hadron
scattering amplitude.
Several models predict an increase in the slope $B(t)$ as $t \rightarrow 0$, which
effectively leads to an additional term
in the  description of  the hadron scattering amplitude.
We shall return to this question later. 

\begin{table*}
\label{tab:1}       %
\begin{center}
\begin{tabular}{lllll}
\noalign{\smallskip}\hline\noalign{\smallskip}
Collaboration & $\sigma_{tot}$ (mb) & $\sigma_{el}/\sigma_{tot}$ & $\rho(t=0)$ & $B(t=0)$ \\
\hline\noalign{\smallskip}
\hline\noalign{\smallskip}
   \cite{BSW} & 103 &  0.28  & 0.12  & 19 \\
    \cite{GLM} & 110.5  & 0.229    &   & 20.5  \\
  \cite{COMPETE} &  111 &    &  0.11 &   \\
   \cite{Paris} & 123.3  &    & 0.103  &  \\
   \cite{SelyuginNP87} & 128  &  0.33 &  0.19  &  21   \\
  \cite{SelyuginBDL06}  &  150 & 0.29 & 0.24  & 21.4 \\
  \cite{Protvino} &  230   & 0.67 &  &  \\
\noalign{\smallskip}\hline
\end{tabular}
\caption{Predictions of different models at $ \ (\sqrt{s} =14 \ $TeV, $t=0$) }
\end{center}
\end{table*}
One should realise that the theoretical predictions are somewhat uncertain. We show in Table 1 recent estimates of the cross section at the LHC.  This is partially due to the fact that the  dispersion of the experimental data for $\sigma_{tot}$
  at high energy above the ISR energies
  is very wide.  We must note that, except for the UA4 and UA4/2 collaborations,
 the other experiments have not published the actual
 numbers for $dN/dt$.
 We can only hope that the new results  from  the LHC experiments
 will not  continue this practice.
 In this context, we must remember the eventual problems that may arise if one fixes $\sigma_{tot}$ or $\rho$ to decrease the size of the errors: indeed, this is what the UA4/2 Collaboration did when they extracted $\rho(0)$, fixing $\sigma_{tot}$
  from  the  UA4 Collaboration ($\sigma_{tot}=61.9 $ mb), or from their own measurement ($\sigma_{tot}=63.0 $ mb). As shown in Table 2, the resulting values for $\rho(0)$ appear inconsistent.
A more careful analysis \cite{SelyuginYF92,SelyuginPL94}  shows
  that there is no contradiction between the measurements of UA4 and UA4/2.

\begin{table*}
\label{tab:2}       
\begin{center}
\begin{tabular}{ll|ll}
\hline\noalign{\smallskip}
 \multicolumn{4}{c} {$\bar{\rho}  \ (\sqrt{s} =540 \ $GeV, $0.000875 \leq |t| \leq 0.12 \ $GeV$^2$)}   \\
\noalign{\smallskip}\hline\noalign{\smallskip}
 experiment      & experimental analysis      & global analysis I \cite{SelyuginYF92} &global analysis II \cite{SelyuginPL94}  \\

 UA4      & $0.24 \pm 0.02$      & $0.19 \pm 0.03$ & - \\
 UA4/2    &  $0.135 \pm 0.015$     & - & $0.17 \pm 0.02$    \\
\noalign{\smallskip}\hline
\end{tabular}
\caption{Average values of $\rho$, derived with fixed total cross section (first two columns), and from a global analysis (last two columns).}
\end{center}
\end{table*}

\section{Fitting procedure for  $\sigma_{tot}(s)$ and Black Disk Limit (BDL)}

   The situation is complicated by the possible transition to the saturation
regime, as the Black Disk Limit (BDL) will be reached at the LHC  \cite{SelyuginBDLCJ04,SelyuginBDL06}. The effect
of  saturation will be a change in the $t$-dependence of $B$ and $\rho$, which will begin for $\sqrt{s} = 2$ to  6  TeV, and which may drastically
change $B(t)$ and $\rho(t)$ at $\sqrt{s} = 14 \ $TeV \cite{SelyuginBDL06}.
As we are about to explain, such a feature can be obtained in very different models.

The first model is based on a fit to soft data which includes a hard pomeron
component \cite{CMSL} of intercept 1.4, which is linked  to the growth of the
gluon density at small $x$ in inelastic processes
\cite{LandshoffHP3}.  
This growth leads to non-linear effects, which saturate the BDL.
Such 
effects were obtained in \cite{SelyuginBDLCJ04,SelyuginBDL06}
and predict that $B(t)$ will increase with $t$ at small $t$ for LHC energies (see
Fig. 1). We also show that the saturation of the BDL will heavily change the
  $t$-dependence of $\rho(t)$, as shown in Fig. 1.
 The hard pomeron component will lead to a decrease of the energy at which
the BDL regime appears, and the effect 
 on the growth of  the total
cross section in uncertain. We show in Fig. 1 and Table 3 the results coming from
an eikonal unitarisation of the amplitude, and we shall refer to this model
\cite{SelyuginBDL06} as
the Eikonalized Soft+Hard Pomeron Model (ESHPM).
\begin{figure}
\includegraphics[width=0.5\textwidth] {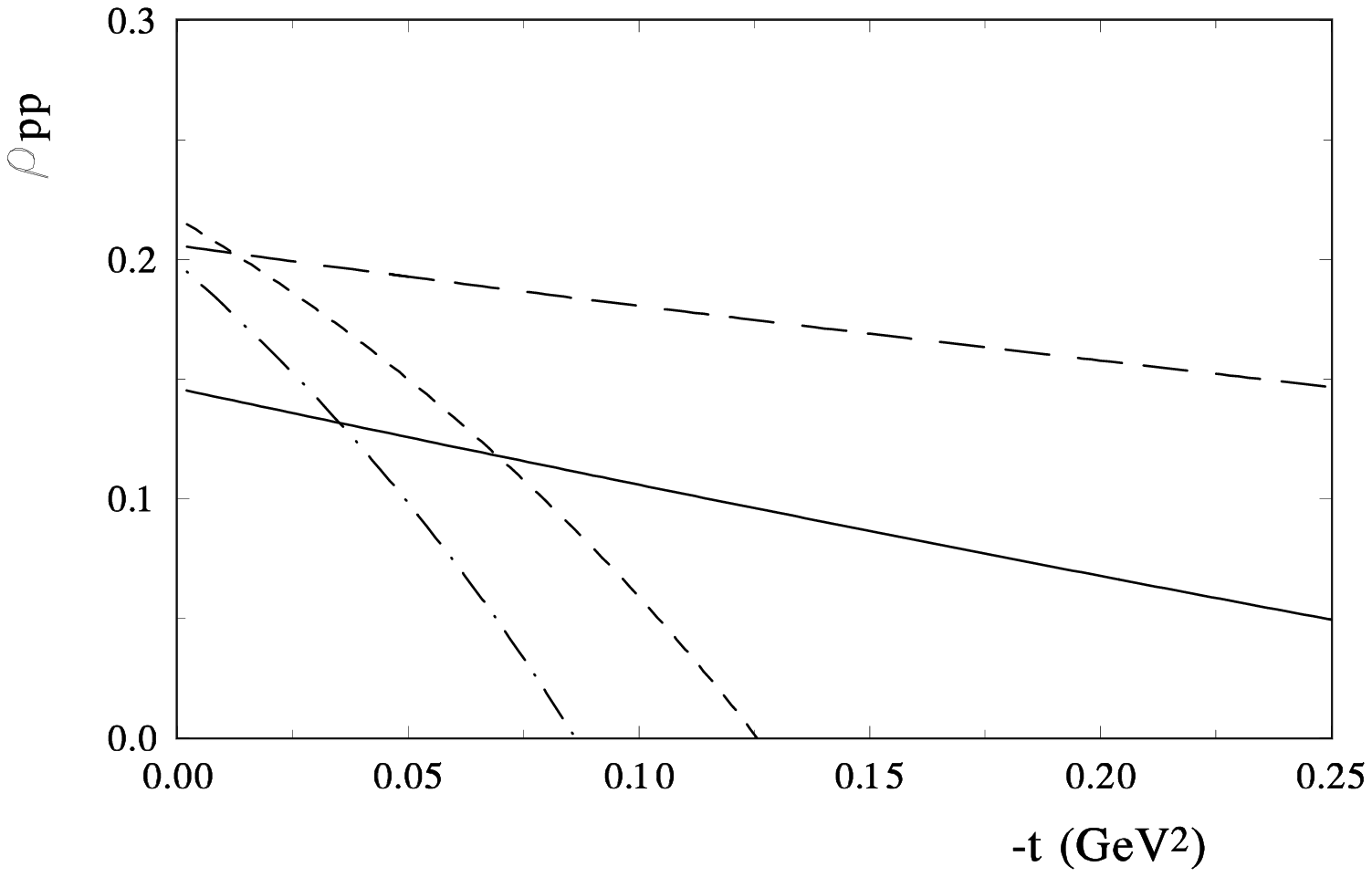}
\includegraphics[width=0.5\textwidth] {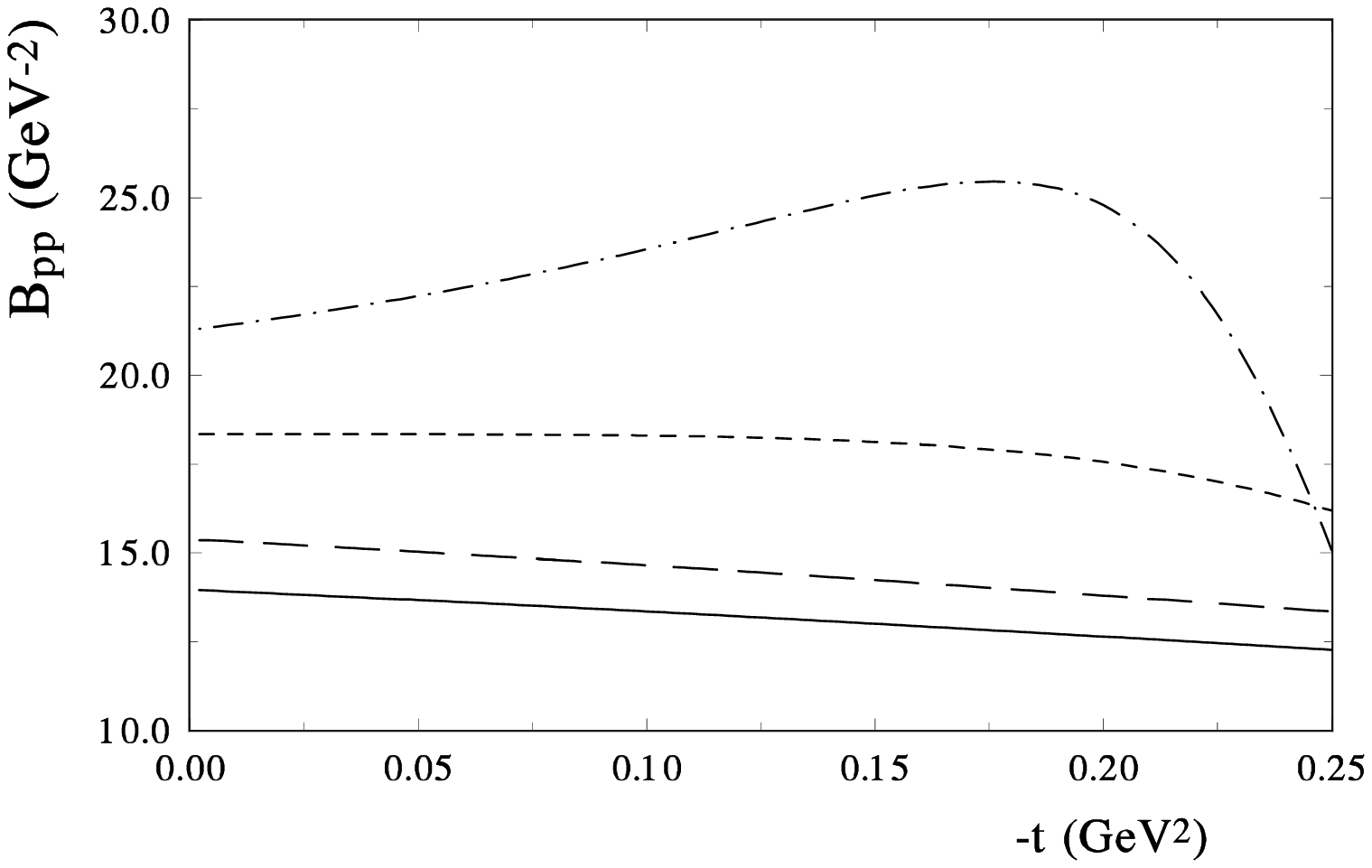}
 \caption{Results of the ESHPM. Left panel:
The ratio of the real to the imaginary part of the amplitude as a function of $t$,
for the bare and the saturated amplitudes
at various energies: 100~GeV (plain curve),
500~GeV (long dashes), 5~TeV (short dashes) and 14~TeV (dash-dotted curve).
Right panel:
The slope of the elastic differential cross section as a function of $t$,
for the bare and saturated amplitudes at various energies: $100$ GeV (plain curve),
$500$ GeV (long dashes), 5 TeV (short dashes) and 14 TeV (dash-dotted curve).}
\end{figure}

\begin{table*}

\label{tab:3}       
\begin{center}
\begin{tabular}{ll|ll}
\hline\noalign{\smallskip}
  \multicolumn{2}{c} {$ t= 0 \ $ } &  \multicolumn{2}{c} {$ t= - 0.1 \ $ GeV $^2$} \\\hline
  {DDM } & ESHPM  & DDM & ESHPM\\\hline
  $0.19 $     & $0.24 $      & $0.08 $ & 0.05 \\
\noalign{\smallskip}\hline
\end{tabular}\caption{Results of the DDM  and of the ESHPM  for $\rho$ at $\sqrt{s} =14$ TeV}
\end{center}
\end{table*}

The second model in which such effects appear is
the Dubna Dynamical model (DDM) of hadron-hadron scattering at high energies
  \cite{SelyuginZC91}. It is based
  on the general principles of quantum field theory
  (analyticity, unitarity and so on) and takes into account basic
  information on the structure of a hadron as a compound system with
  a central region in which the valence quarks are concentrated, and
  a long-distance region filled with a color-singlet quark-gluon field.
  As a result, the hadron amplitude can be  represented  as  a
  sum  of a central  and a peripheral part.
  The DDM predicts that the interaction of the Pomeron with the meson cloud of the
hadrons will give an additional term growing like $\sqrt{s}$.
  This term will become important for energies $\sqrt s \ge 500$ GeV.
This peripheric effect will lead to a saturation of the
  overlapping function $G(b)$, see Fig. 2.
 At small momentum transfer, the DDM predictions agree with the
  experimental data at $\sqrt{s} = 1.8 \ $ TeV. Interestingly, as shown in Fig. 2,
 the DDM predicts that the differential cross sections
 at $-t \approx 0.3 \ $ GeV$^2$ will coincide for all high energies.
 Here again, the $t$-dependence of the slope-$B(s,t)$ will change its behavior
 at LHC energies because of saturation effects.

\begin{figure}
\includegraphics[width=0.5\textwidth] {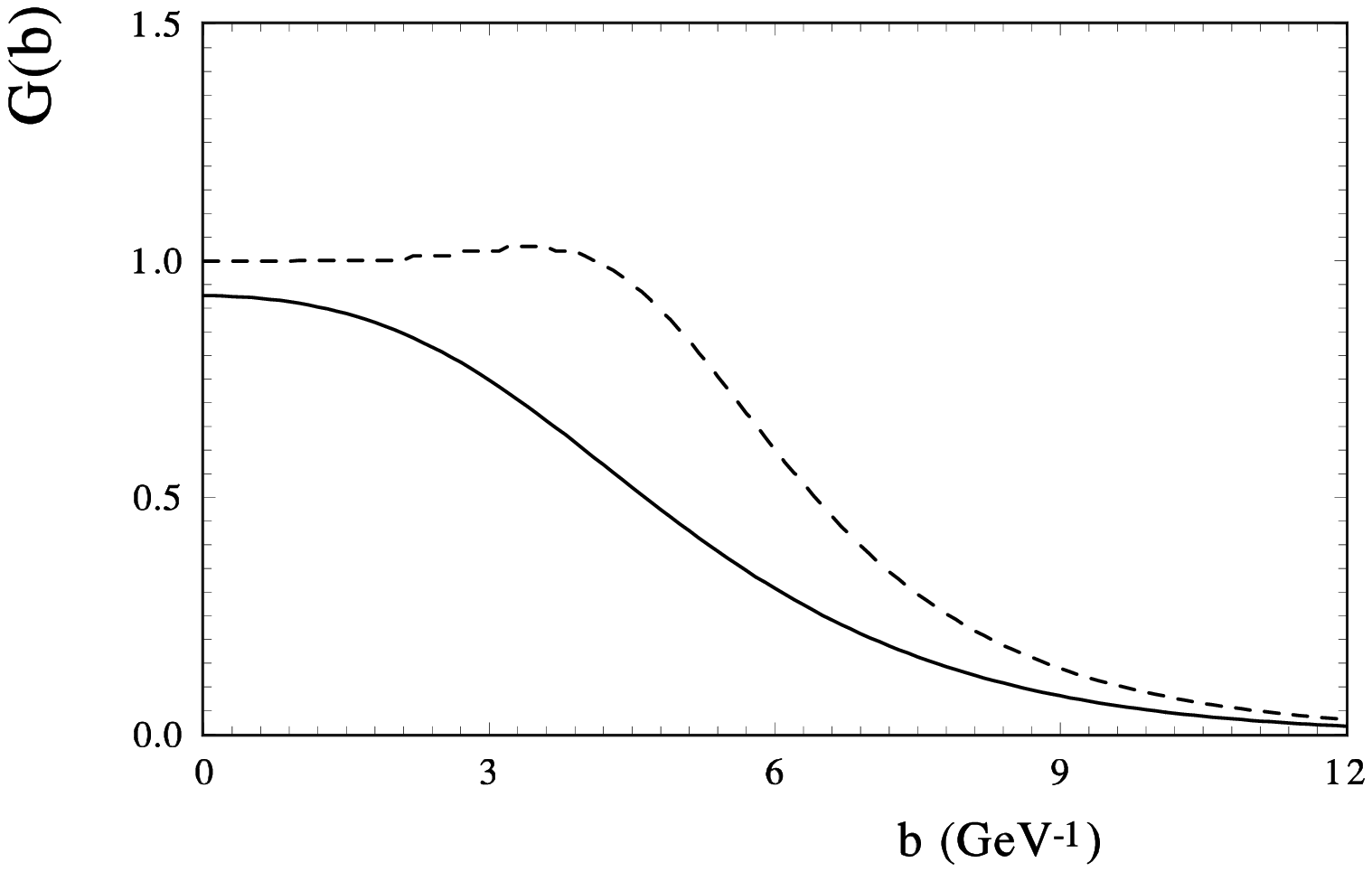}
\includegraphics[width=0.5\textwidth] {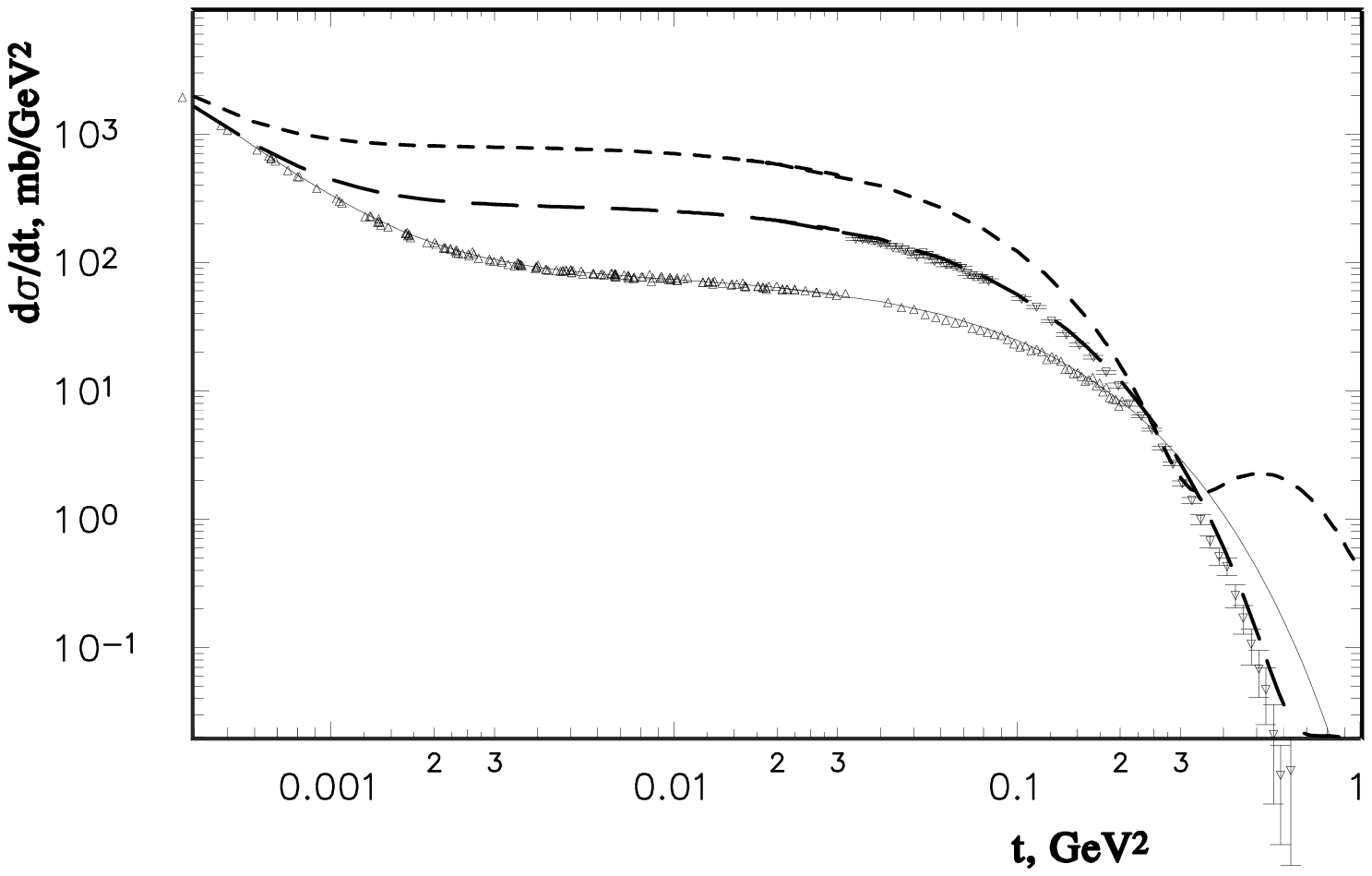}
\caption{ Predictions of the DDM. 
Left panel: Overlapping function at $\sqrt{s} = 2 \ $TeV (solid line) and
at $\sqrt{s} = 14 \ $TeV (dashed line).
Right panel:
 Differential cross sections  at $\sqrt{s} = 23.4 \ $GeV (solid thin line) and
at $\sqrt{s} = 1.8 \ $TeV (dashed line) and at $\sqrt{s} = 14 \ $TeV (solid thick line).
  }
\end{figure}

    Let us now examine what the standard fitting procedure might give at the LHC in the case of saturation of the BDL, which leads to a behaviour of the scattering amplitude very far from an exponential. As an input, we shall use the predictions
for the differential cross sections in the framework of the DDM  for two energies
 $\sqrt{s} =2 \ $TeV and   $\sqrt{s} =14 \ $TeV. For the first energy, the deviation from an exponential is small, whereas it becomes essential at the LHC.
We can simulate the future experimental data from this theoretical
differential cross sections and assume that  $90$ points will be measured in a $t$ interval identical to that of the UA4/2 experiment.
We then randomize the theoretical curve assuming Gaussian errors similar to those of UA4/2. After that, we can fit the simulated data with an exponential amplitude.
The results of this exercise are shown in Tables 4 and 5.
It is clear that at $\sqrt{s} = 14 \ $TeV, the simulated data differ significantly from the results of the fit, especially if one allows for a refitting of their normalisation.

\begin{table*}

\label{tab:4}       %
\begin{center}
\begin{tabular}{lllll}
\hline\noalign{\smallskip}
  $ \sum \chi^{2}_{i} $ & $\sigma_{tot} (mb)$ &  $\rho(t=0)$ &  B(0) (GeV$^{-2}$)& normalization coefficient \\
\hline\noalign{\smallskip}
$91.2$ & $ 82.3 \pm  0.3 $ & $0.15_{fixed}$ & $18.1 \pm 0.2$ & $1_{fixed}$  \\
$88.3$ & $ 85. \pm  1.7 $ & $0.15_{fixed}$ & $18.16 \pm 0.2$ & $ 0.94 \pm 0.04$  \\
$89.3$ & $ 82.3 \pm  0.3 $ & $0.18 \pm 0.02$ & $18.3 \pm 0.2$ & $1_{fixed}$  \\
$88.1$ & $ 85.2 \pm  3. $ & $0.147 \pm 0.04$ & $18.1 \pm 0.25$ & $ 0.93 \pm 0.07$  \\
\noalign{\smallskip}\hline
\end{tabular}
\caption{Fits  at $\sqrt{s} = 2 \ $TeV [Input $\rho(0) =0.23$; $\  \  \sigma_{tot}= 82.7 \ $mb;
 $B(0)=18.3 \ $GeV$^{-2}$  )].}\end{center}
\end{table*}

\begin{table*}

\label{tab:5}       %
\begin{center}\begin{tabular}{lllll}
\hline\noalign{\smallskip}
 $  \sum \chi^{2}_{i} $ &    $\sigma_{tot} $ (mb)   & $\rho(t=0)$ &  B(0) (GeV$^{-2}$)
& normalization coefficient \\
\hline\noalign{\smallskip}
$133$ & $ 155.3 \pm  0.5 $ & $0.15_{fixed}$ & $23.1 \pm 0.2$ & $1_{fixed}$  \\
$120$ & $ 180. \pm  8.6 $ & $0.15_{fixed}$ & $23.2 \pm 0.15$ & $ 0.74 \pm 0.07$  \\
$109$ & $ 153.4 \pm  0.7 $ & $0.26 \pm 0.03$ & $23.5 \pm 0.17$ & $1_{fixed}$  \\
$108$ & $ 142.3 \pm  2.8 $ & $0.29 \pm 0.05$ & $23.6 \pm 0.2$ & $ 1.15 \pm 0.05$  \\
\noalign{\smallskip}\hline
\end{tabular}
\caption{Fits  at $\sqrt{s} = 14 \ $TeV
[Input $\rho(0) =0.24$; $\   \sigma_{tot}= 152.5 \ $mb; $B(0)=21.4 \ $GeV$^{-2}$  ].}
\end{center}\end{table*}

Saturation of the profile function will surely control
the behaviour of $\sigma_{tot}$ at higher
energies and will result in a significant decrease of the LHC cross section.
However, it is clear that the simple saturation
considered here is not enough, as the total cross section at the Tevatron will
be 85 mb, which is 2 standard deviations from the CDF result.
However, the increase of the slope with $t$ at small $t$ is a generic feature
of all saturating models.

\section{Oscillations and additional method}

      As the standard fitting procedure can give misleading results, we need to find an
      additional method to define or check the basic parameters
      of the hadron scattering amplitude.
      Especially as there can be additional specific features in the $t$-dependence
      of the different parts of the amplitude.
      For example, there can be some oscillations in the differential cross sections
    which can come from different sources.
   It was shown  \cite{AKM} that
   if the Pomeranchuk
   theorem is broken and the scattering amplitude
   grows to a maximal possible extent,
the elastic scattering cross section would exhibit
a periodic structure in $q = \sqrt{|t|}$
at small $-t$.  It was shown \cite{Selyugin96} that the oscilations in the UA4/2 data
over $\sqrt{|t|}$ can be connected with a rigid-string potential
or with residual long-range forces between nucleons.
These small oscillations in the differential cross section are difficult to detect
 by the standard fitting method. Another method was proposed, which consists
in the comparison of two statistically
independent samples built by binning the whole $t$-interval in small intervals,
 proportional to $\sqrt{|t|}$, and by keeping one interval out of two.
The deviations of the
experimental values from theoretical expectations, weighted by the experimental error,
 are then summed for each sample $k$:
\cite{SelyuginOSC}.
\begin{equation}
\Delta R^k(t)=\sum_{|t_i|<|t|} \Delta R^k_ {i} = \sum_{|t_i|<|t|}[(d\sigma^k/dt_{i})^{exp}
  -  (d\sigma/dt_{i})^{th}] / \delta_{i}^{exp},
\end{equation}
 where $\delta_{i}^{exp} $ is the experimental error.
   This method gives two curves which statistically coincide if oscillations are absent and which grow apart  with $t$ if the oscillations are present.

If the theoretical
 curve does not precisely describe the experimental data,
 (for example, if the physical
 hadron amplitude does not have an exactly
 exponential behavior with momentum transfer), the sum $\Delta R^k(t) $
 will differ from zero, going beyond the size
 of a statistical error. This method thus gives the possibility to check the
validity of the model assumptions and of the
 parameters which describe the hadron scattering amplitude.
Note that another  specific method was proposed in \cite{SelyuginBL95,SelyuginPL05}.

\section{Conclusion}
   As the cross section of proton elastic scattering will be measured at the LHC,
we need to know more about the behaviour of
  the hadron scattering amplitude at small $t$.
  The analysis of soft data, taking into account the integral dispersion relations,
 shows a contribution of the hard pomeron in elastic scattering.
  In this case, it is very likely that
   at the LHC we shall reach the saturation regime called the BDL 
   It will manifest itself
  in the behavior of $B(t) $ and of $ \rho(t) $ and lead to a non-exponential behavior
  of the hadron scattering amplitude at small $t$, which will depend
  on the form of the unitarization procedure.
   In other words, different impact parameter dependences of the scattering amplitude
   will  lead to different energy dependences of the ratio of the elastic
   to the total cross sections.

The regime of the BDL may correspond to parton saturation in the interacting hadrons,
 which is described by a non-linear equation.
Indeed, there is a one-to-one correspondance between non-linear equations
  and the different forms of the unitarization schemes.

The possibility of a new behaviour of $\rho(s,t)$ and $B(s,t)$ at LHC energies
 has to be taken into account in
the procedure extracting the value of the total cross sections by the standard
fitting method. It is needed to use additional specific methods for the determination
of the size of the total cross section and of $\rho(s,t)$,
 such as calculating $\Delta R$ and comparing independent choices.

\vspace{0.5cm}

{\small The authors would like to thank  for helpful discussions E. Martynov and P.V. Landshoff.
 O.S. gratefully acknowledges financial support
  from FRNS and would like to thank the  University of Li\`{e}ge
  where part of this work was done.
    }

\begin{footnotesize}
\bibliographystyle{blois07}
{\raggedright
\bibliography{selyugin}
}

\end{footnotesize}

\end{document}